\documentclass[final]{cvpr}
\usepackage{times}
\usepackage{epsfig}
\usepackage{graphicx}
\usepackage{amsmath}
\usepackage{amssymb}
\usepackage{comment}
\usepackage{makecell}
\usepackage{soul}
\usepackage{multirow}

\usepackage[pagebackref=true,breaklinks=true,colorlinks,bookmarks=false]{hyperref}

\usepackage{glossaries}
\usepackage[table]{xcolor}

\usepackage{footnote}
\makesavenoteenv{tabular}
\makesavenoteenv{table}

\definecolor{franzcolor}{rgb}{0.1, 0.8, 0.4}
\definecolor{darkocolor}{rgb}{0.1, 0.4, 0.8}
\definecolor{christiancolor}{rgb}{0.8, 0.4, 0.1}
\definecolor{redcolor}{rgb}{0.8, 0.0, 0.0}

\newcommand{\lightvalue}{0.92}
\definecolor{lightgreen}{rgb}{\lightvalue,1,\lightvalue}
\definecolor{lightred}{rgb}{1,\lightvalue,\lightvalue}
\definecolor{lightblue}{rgb}{\lightvalue,\lightvalue,1}
\definecolor{lightyellow}{rgb}{1,1,\lightvalue}

\newcommand{\imagelettersmall}{x}
\newcommand{\groundtruthlettersmall}{y}

\newcommand{\image}{\mathbf{\imagelettersmall}}
\newcommand{\groundtruth}{\mathbf{\groundtruthlettersmall}}

\newcommand{\prediction}{\mathbf{\hat{\groundtruthlettersmall}}}

\newcommand{\lossbase}{L}
\newcommand{\lossdice}{\lossbase_{\text{GD}}}
\newcommand{\lossce}{\lossbase_{\text{CE}}}

\newcommand{\lossfactorbase}{\lambda}

\newcommand{\letterlocalization}{\text{loc}}
\newcommand{\lettersegmentation}{\text{seg}}

\newcommand{\modelbase}{\mathcal{M}}
\newcommand{\modellocalization}{\modelbase_{\letterlocalization}}
\newcommand{\modelsegmentation}{\modelbase_{\lettersegmentation}}

\newcommand{\losslocalization}{\lossbase_{\letterlocalization}}
\newcommand{\losssegmentation}{\lossbase_{\lettersegmentation}}

\newcommand{\inputlocalization}{\image_{\letterlocalization}}
\newcommand{\inputsegmentation}{\image_{\lettersegmentation}}
\newcommand{\groundtruthlocalization}{\groundtruth_{\letterlocalization}}
\newcommand{\groundtruthsegmentation}{\groundtruth_{\lettersegmentation}}
\newcommand{\predictionlocalization}{\prediction_{\letterlocalization}}
\newcommand{\predictionsegmentation}{\prediction_{\lettersegmentation}}

\newcommand{\letterlocal}{1}
\newcommand{\letterspatial}{2}

\newcommand{\inputsegmentationlocal}{\image_{\lettersegmentation}^{\letterlocal}}
\newcommand{\inputsegmentationspatial}{\image_{\lettersegmentation}^{\letterspatial}}
\newcommand{\predictionsegmentationlocal}{\prediction_{\lettersegmentation}^{\letterlocal}}
\newcommand{\predictionsegmentationspatial}{\prediction_{\lettersegmentation}^{\letterspatial}}

\newcommand{\lossfactorsegmentationlocal}{\lossfactorbase_{\lettersegmentation}^{\letterlocal}}
\newcommand{\lossfactorsegmentationspatial}{\lossfactorbase_{\lettersegmentation}^{\letterspatial}}

\newacronym[plural=SCNs,firstplural=SpatialConfiguration-Nets (SCN)]{scn}{SCN}{SpatialConfiguration-Net}
\newacronym[plural=CNNs,firstplural=Convolutional Neural Networks (CNN)]{cnn}{CNN}{Convolutional Neural Network}
\newacronym[plural=ASMs,firstplural=Active Shape Models (ASM)]{asm}{ASM}{Active Shape Model}

\newacronym{roi}{ROI}{region of interest}
\newacronym{flops}{FLOPs}{Floating Point Operations}

\newacronym{ct}{CT}{Computed Tomography}
\newacronym{mr}{MR}{Magnetic Resonance}

\newacronym{dsc}{DSC}{Dice Similarity Coefficient}
\newacronym{nsd}{NSD}{Normalized Surface Distance}

\newacronym[plural=GPUs,firstplural=Graphics Processing Units (GPU)]{gpu}{GPU}{Graphics Processing Unit}

\def\confYear{MICCAI FLARE 2021}

\begin{document}

\title{Efficient Multi-Organ Segmentation Using SpatialConfiguration-Net \\ with Low GPU Memory Requirements}

\author{Franz Thaler\\
Gottfried Schatz Research Center: Biophysics\\
Medical University of Graz, Austria\\
and\\
Institute of Computer Graphics and Vision\\
Graz University of Technology, Austria\\
{\tt\small franz.thaler@medunigraz.at}
\and
Christian Payer\\
Institute of Computer Graphics and Vision\\
Graz University of Technology, Austria\\
{\tt\small christian.payer@icg.tugraz.at}
\and
Horst Bischof\\
Institute of Computer Graphics and Vision\\
Graz University of Technology, Austria\\
{\tt\small bischof@icg.tugraz.at}
\and
Darko \v{S}tern\\
Gottfried Schatz Research Center: Biophysics\\
Medical University of Graz, Austria\\
{\tt\small darko.stern@medunigraz.at}
}

\maketitle

\begin{abstract}

Even though many semantic segmentation methods exist that are able to perform well on many medical datasets, they are often not designed for direct use in clinical practice.
The two main concerns are generalization to unseen data with a different visual appearance, e.g., images acquired using a different scanner, and efficiency in terms of computation time and required Graphics Processing Unit (GPU) memory.
In this work, we employ a multi-organ segmentation model based on the SpatialConfiguration-Net (SCN), which integrates prior knowledge of the spatial configuration among the labelled organs to resolve spurious responses in the network outputs.
Furthermore, we modified the architecture of the segmentation model to reduce its memory footprint as much as possible without drastically impacting the quality of the predictions.
Lastly, we implemented a minimal inference script for which we optimized both execution time and required GPU memory.

\end{abstract}

\section{Introduction}

The invention of medical imaging techniques like X-ray imaging, \gls{ct} and \gls{mr} imaging had a huge impact on clinical workflows and greatly improved patient treatment by allowing non-invasive visualization of the patients' anatomy.
Semantic segmentation in medical imaging is an important tool in clinical practice and is used in radiotherapy to accurately delineate tumors and treat certain cancers~\cite{liao2018bayesian}, for morphological analysis of organs to infer information like the volume and shape of, e.g., the liver~\cite{gotra2017liver} or for surgery planning~\cite{virzi2020comprehensive}.
In medical imaging, an expert is required to create a manual semantic segmentation, which is not only laborious and time-consuming, but also cost-intensive and consequently not always feasible in clinical practice.
This leads to an increased demand for automated methods for medical image segmentation, where only minimal interaction by an expert is necessary to generate a segmentation.

Modern machine learning methods for medical image segmentation are based on \glspl{cnn} for which a large number of publications can already be found in the literature, ranging from multi-organ segmentation~\cite{fu2021review} and cardiac image segmentation~\cite{chen2020deep} to brain tumor segmentation~\cite{tiwari2020brain}.
Many approaches are based on variants of the well-known U-Net architecture~\cite{ronneberger2015u}, which was shown to perform well on many datasets.
However, many datasets consist only of images acquired from a single center or scanner, or only include cases with a single disease, and thus the generalization property of models trained on such datasets is often not evaluated.
Furthermore, most approaches only aim to maximize the quality of the predictions without considering model efficiency, which often leads to an increased complexity of the model and consequently reduces the model's applicability in clinical practice.
Even though standard semantic segmentation methods like the U-Net~\cite{ronneberger2015u} perform well on many datasets, closer inspection of the predictions for unseen images reveals that some spurious responses remain even after the network has been fully trained.
This becomes even more apparent when the images differ from the training data, e.g., if they have been acquired with a different scanner from which no images have been used to train the model, or when the complexity of the model is reduced to increase its applicability in practice.

\begin{figure*}[htbp]
\centering
\includegraphics[width=1.0\textwidth]{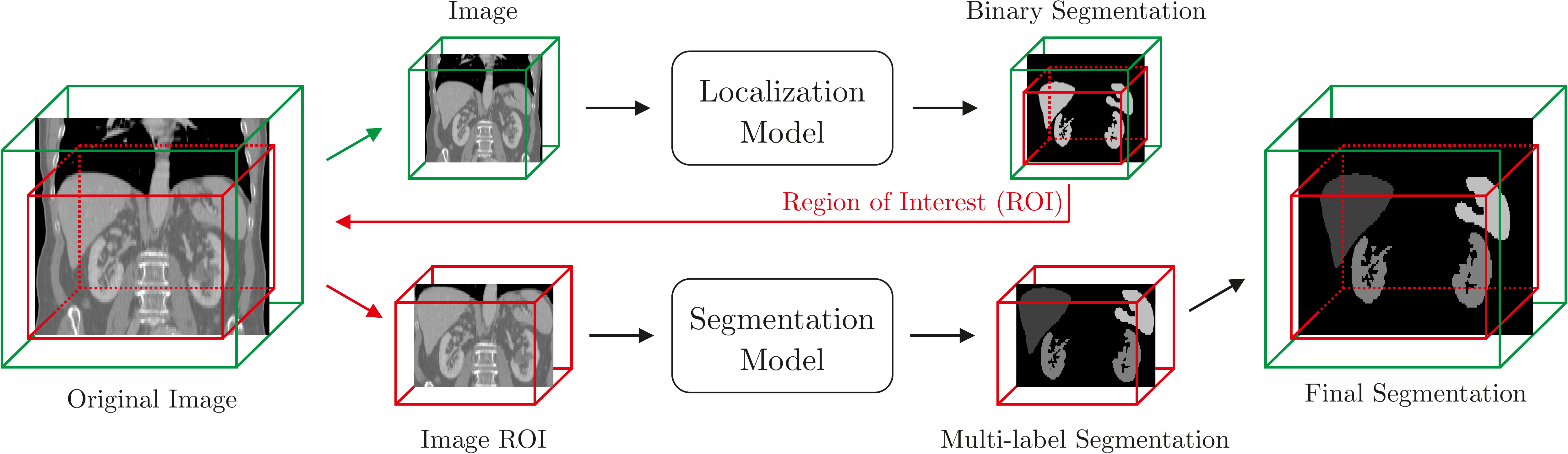}
\caption{
Schematic overview of our method.
First, using the full image content at a low resolution, we apply a localization model to predict a coarse binary segmentation from which we extract the region of interest (ROI) for further use.
Then, the ROI is extracted from the original image at a high resolution and used as the input of our segmentation model to predict the final multi-label segmentation.
Lastly, the final segmentation is resampled to match the original image.
}
\label{fig:overview}
\end{figure*}

Motivated to improve the quality of the predictions for unseen images without using any additional data, we propose to use the \gls{scn}, which was originally applied for landmark localization~\cite{payer2016regressing,payer2019integrating} and later adapted to multi-label segmentation of the heart~\cite{payer2017multi}. 
We hypothesize that the \gls{scn}, which consists of a local and a spatial network, results in better predictions for unseen images as the spatial network integrates prior knowledge of the spatial configuration among the labelled organs, which can be used to resolve spurious responses.
Furthermore, we carefully modified the \gls{scn} to reduce model complexity as much as possible without drastically impacting the quality of the segmentation results by observing the trade-off between efficiency and segmentation quality.
Lastly, we optimized the code to generate predictions for unseen images by minimizing computation time as well as the required \gls{gpu} memory.

\section{Method}

In this work, we propose a multi-organ segmentation approach, where first a localization model is used to determine the \gls{roi} before employing a segmentation model based on the \gls{scn}, see Fig.~\ref{fig:overview}.
Our code is available on GitHub\footnote{https://github.com/franzthaler/FLARE21-EfficientSCN}.

\subsection{Preprocessing}

\paragraph{Localization of the Region of Interest}

The utilized localization model receives the full volumetric image content as input.
As the intensity values of \gls{ct} images are well defined by the Hounsfield scale, we do not perform an intensity normalization on a per-image basis.
However, we divide the intensity values of all our images by 2048 and clip them to $[-1,1]$.
After performing a Gaussian smoothing with $\sigma = 3$, we linearly resample the images to a spacing of $6 \times 6 \times 6$ mm and an image size between $32 \times 32 \times 32$ and $80 \times 80 \times 256$ depending on the physical resolution of the original image, yielding $\inputlocalization$.
In cases where the resampled images would still not fit the maximum image size, we increase the image spacing further such that the resampled image will fit.

\begin{table*}[!htbp]
\caption{Data split of FLARE 2021.}
\label{tab:dataset}
\centering
\begin{tabular}{llll}
\hline
Data Split
& Center& Phase & \# Num.\\
\hline
\multirow{2}{*}{Training ( 361 cases )}& The National Institutes of Health Clinical Center & portal venous phase & 80 \\
 & Memorial Sloan Kettering Cancer Center& portal venous phase & 281\\
\hline
\multirow{3}{*}{Validation ( 50 cases )}& Memorial Sloan Kettering Cancer Center& portal venous phase & 5\\
 & University of Minnesota & late arterial phase & 25 \\
 & 7 Medical Centers & various phases& 20 \\
\hline
\multirow{4}{*}{Testing ( 100 cases )}& Memorial Sloan Kettering Cancer Center& portal venous phase & 5\\
 & University of Minnesota & late arterial phase & 25 \\
 & 7 Medical Centers& various phases& 20 \\
 & Nanjing University& various phases& 50 \\
\hline
\end{tabular}
\end{table*}

\paragraph{Multi-Organ Segmentation}

Our segmentation model utilizes the \gls{roi} predicted by the localization model to reduce the memory footprint of the image by cropping it. This allows the content within the \gls{roi} to retain a higher resolution compared to the image used by our localization model.
Consequently, all images are cropped to a size between $32 \times 32 \times 32$ and $160 \times 128 \times 160$ with a spacing of $2 \times 2 \times 2$ mm or larger, in case the \gls{roi} would exceed the maximum image size.
To train our segmentation model, we computed the bounding box containing every labelled voxel directly from the ground truth for each training sample.
During inference, where the ground truth label is not available, we used the \gls{roi} predicted by our localization model to retrieve the relevant region of the image.
In both cases, we additionally enlarged the \gls{roi} by up to $16$ voxels in both directions of each dimension independently, depending on the remaining distance to the respective image border, to also include some of the surrounding anatomy, yielding the input of our segmentation model $\inputsegmentation$.
Again, the intensity values of each image were divided by 2048 and clipped to $[-1,1]$.

\subsection{Proposed Method}

\paragraph{Localization of the Region of Interest}

Our localization model $\modellocalization$ is implemented as a coarse binary segmentation \gls{cnn} where all labelled organs are treated as the foreground label and which is trained using the full image content after greatly downsampling the image to reduce the memory footprint.
First, the image $\inputlocalization$ is used as input for the localization model to predict a binary segmentation $\predictionlocalization = \modellocalization(\inputlocalization)$.
The binary segmentation is then used to compute a bounding box containing every segmented voxel, which we define as the \gls{roi}.
The network architecture of the localization model was adapted from the U-Net~\cite{ronneberger2015u} and trained using a generalized Dice loss~\cite{sudre2017generalised}, i.e.,
\begin{equation}
\begin{aligned}
\losslocalization = \lossdice(\groundtruthlocalization, \predictionlocalization),
\end{aligned}
\end{equation}
where $\groundtruthlocalization$ represents the ground truth label resampled to the same size and spacing as the image $\inputlocalization$.
Our localization model uses a total of 637,474 parameters and the total number of \gls{flops} for one forward pass ranges from 8,613,207,612 for input size $32 \times 32 \times 32$ to 430,660,377,660 for input size $80 \times 80 \times 256$.

\paragraph{Multi-Organ Segmentation}

The proposed segmentation model $\modelsegmentation$ uses the image $\inputsegmentation$ as input and is trained to predict a multi-label segmentation $\predictionsegmentation$, i.e., $\predictionsegmentation = \modelsegmentation(\inputsegmentation)$.
The architecture of our segmentation model is based on the \gls{scn}, which was originally proposed for landmark localization~\cite{payer2016regressing,payer2019integrating} and later adapted to multi-label segmentation of the whole heart~\cite{payer2017multi}. 
In this work, we adapted the \gls{scn} to multi-organ segmentation and simultaneously reduced its memory footprint.
To train the proposed architecture, we employed a supervised loss on the final prediction $\predictionsegmentation$ of the segmentation model as well as on the intermediate predictions of the local and spatial networks to ensure that both subnetworks lead to a plausible prediction on their own.
Defining the input and output of the local network as $\inputsegmentationlocal$ and $\predictionsegmentationlocal$ and of the spatial network as $\inputsegmentationspatial$ and $\predictionsegmentationspatial$ allows us to formulate the loss function as
\begin{equation}
\begin{aligned}
\losssegmentation 
&= \lossdice(\groundtruthsegmentation, \predictionsegmentation)
+ \lossfactorsegmentationlocal \lossce(\groundtruthsegmentation, \predictionsegmentationlocal) \\
&+ \lossfactorsegmentationspatial \lossce(\groundtruthsegmentation, \predictionsegmentationspatial),
\end{aligned}
\end{equation}
where the ground truth $\groundtruthsegmentation$ was resampled identically to the image $\inputsegmentation$.
The terms $\lossdice$ and $\lossce$ represent the generalized Dice loss and the cross-entropy loss, respectively, and $\lossfactorsegmentationlocal$ and $\lossfactorsegmentationspatial$ are hyperparameters used to weight the individual loss functions.
Overall, our segmentation model uses 1,270,090 parameters and requires 8,797,627,020 \gls{flops} for an input size of $32 \times 32 \times 32$ and 879,762,672,300 \gls{flops} when using an input size of $160 \times 128 \times 160$.

\subsection{Post-processing}

We do not perform any post-processing other than resampling the final prediction to the same size, spacing, etc. as the original input.

\section{Dataset and Evaluation Metrics}

\subsection{Dataset}

The dataset used in the FLARE 2021 Challenge is adapted from MSD~\cite{simpson2019MSD} (Liver~\cite{bilic2019lits}, Spleen, Pancreas), NIH Pancreas~\cite{roth9data,roth2015deeporgan,clark2013cancer}, KiTS~\cite{KiTS,KiTSDataset}, and Nanjing University under the license permission.
For more detailed information on the dataset, please refer to the challenge website and~\cite{AbdomenCT-1K}.

In total, the dataset consists of 511 images.
The dataset is split into a training, validation and test set containing approximately $70\%$, $10\%$ and $20\%$ respectively, resulting in 361 training cases, 50 validation cases, and 100 test cases.
More detailed information is presented in Table~\ref{tab:dataset}.

\subsection{Evaluation Metrics} 

In the evaluation, both accuracy and efficiency of the model will be considered to rank the challenge participants using the following metrics:
\begin{itemize}
    \item Dice Similarity Coefficient (DSC)
    \item Normalized Surface Distance (NSD)
    \item Running time
    \item Maximum used GPU memory (when the inference is stable)
\end{itemize}

\section{Implementation Details}

\subsection{Environments and requirements}

We implemented our method in Python and used TensorFlow as the deep learning framework.
SimpleITK was used to read and write the medical images as well as their meta-data to retrieve the image information as NumPy arrays.
Detailed information on the environment used during the development of our method is given in Table~\ref{table:env}.

\begin{table}[!htbp]
\caption{Environment and requirements.}
\label{table:env}
\begin{center}
\resizebox{0.47\textwidth}{!}{
\begin{tabular}{m{3.2cm}<\raggedright|m{5.3cm}<\raggedright} 
\hline
Operating System & Linux Manjaro 5.13.5-1 \\
\hline
CPU & AMD Ryzen 9 3900X \\
\hline
RAM & 16$\times$4GB \\
\hline
GPU & NVIDIA GeForce RTX 2080 Ti \\
\hline
CUDA version & 11.4 \\
\hline
Programming language & Python 3.9 \\ 
\hline
Deep learning framework & TensorFlow 2.6 \\
\hline
Specification of dependencies & SimpleITK 2.0, NumPy 1.20, SciPy 1.7, scikit-image 0.18, tqdm 4.61 \\
\hline
\end{tabular}
}
\end{center}
\end{table}

\subsection{Training protocols}

While the localization model is based on the U-Net~\cite{ronneberger2015u} and the segmentation model on the \gls{scn}~\cite{payer2017multi}, the architecture of the localization model as well as the architectures of the local and spatial networks of the segmentation model follow the same structure.
Namely, each architecture consists of a sequence of contracting blocks that are followed by the same number of expanding blocks.
Additionally, each contracting block is connected by a skip connection to the expanding block on the same level of depth, similar to the U-Net~\cite{ronneberger2015u}.
Each block consists of two consecutive convolution and dropout layers \cite{hinton2012improving}.
Furthermore, each convolution, with the exception of the final one, employs a kernel size of $3 \times 3 \times 3$ and leaky ReLU~\cite{he2015delving} with $\alpha = 0.1$.
The final convolution uses a linear activation function and the number of filters is set to the number of labels.
The dropout ratio is set to 0.1 and we use average pooling and linear upsampling with a stride of $2 \times 2 \times 2$ as well as He initialization \cite{he2015delving}.
We use a mini-batch size of 1, employ Adam \cite{kingma2014adam} as the optimizer with a learning rate of 0.0001 and compute the model weights as an exponential moving average.
Our localization model as well as the local network of our segmentation model use 5 levels of depth and 32 filters for intermediate convolution layers, while the spatial network of our segmentation model uses 4 levels of depth and 16 filters.
Whenever a sample is selected during training, we perform data augmentation via a random transformation consisting of a translation, rotation, scaling and elastic deformation, as well as an intensity shift and scale.
See Table~\ref{table:training} for more details.

To develop our method and evaluate the performance of our segmentation model, we conducted a 4-fold cross-validation for which we split the 361 labelled images of the training set into 4 random folds.
Each segmentation model in our cross-validation experiment was trained for 100,000 iterations.
During our experiments, we observed that a model with reduced complexity is able to catch up to the performance of a more complex model when it is trained for more iterations.
Thus, for our final submission, we increased the number of iterations to train the localization and segmentation models to 1,000,000 and 500,000, respectively.

\begin{table}[!htbp]
\caption{Training protocols.}
\label{table:training}
\begin{center}
\resizebox{0.47\textwidth}{!}{
\begin{tabular}{m{3.2cm}<\raggedright|m{5.3cm}<\raggedright} 
\hline
Data augmentation methods & Rotation, translation, scaling, elastic deformation, intensity shift and scaling \\
\hline
Initialization of the network & He normal initialization \\
\hline
Batch size& 1 \\
\hline
Patch sampling strategy $\modellocalization$ & full image \\
\hline
Patch size $\modellocalization$ & [$32 \times 32 \times 32$] to [$80 \times 80 \times 256$] \\
\hline
Image spacing $\modellocalization$ & $6 \times 6 \times 6$ mm \\ 
\hline
Total iterations $\modellocalization$ & 1,000,000 \\
\hline
Patch sampling strategy $\modelsegmentation$ & \gls{roi} from ground truth (training) or localization model (inference) \\
\hline 
Patch size $\modelsegmentation$ & [$32 \times 32 \times 32$] to [$160 \times 128 \times 160$] \\ 
\hline
Image spacing $\modelsegmentation$ & $2 \times 2 \times 2$ mm \\ 
\hline
Total iterations $\modelsegmentation$ & 500,000 \\
\hline
Optimizer& Adam ($\beta_1=0.9$, $\beta_2=0.999$)\\ \hline
Initial learning rate& 0.0001 \\ \hline
Learning rate decay schedule & constant \\
\hline
Stopping criteria, and optimal model selection criteria & Stopping criterion is reaching the maximum number of iterations. \\ 
\hline
Training time $\modellocalization$ & 11.0 hours \\
\hline
Training time $\modelsegmentation$ & 49.0 hours \\
\hline 
CO$_2$eq\footnote{https://github.com/lfwa/carbontracker/} $\modellocalization$ & 1,183 g \\
\hline 
CO$_2$eq $\modelsegmentation$ & 7,136 g \\
\hline
\end{tabular}
}
\end{center}
\end{table}

\begin{table*}[htbp]
\normalsize
\centering
\caption{Quantitative results of 4-fold cross-validation in terms of DSC and NSD.}
\label{tab:cross-validation}
\renewcommand\tabcolsep{3pt}
\begin{tabular}{cccccccccc}
\hline
\multirow{2}{*}{Training} & \multicolumn{2}{c}{\cellcolor{lightred} Liver} & \multicolumn{2}{c}{\cellcolor{lightgreen} Kidney} & \multicolumn{2}{c}{\cellcolor{lightblue} Spleen} & \multicolumn{2}{c}{\cellcolor{lightyellow} Pancreas} \\
\cline{2-9} & DSC (\%) & NSD (\%) & DSC (\%) & NSD (\%) & DSC (\%) & NSD (\%) & DSC (\%) & NSD (\%) \\
\hline
Fold-1 & $97.6 \pm 2.1$ & $88.4 \pm 6.5$ & $95.0 \pm 3.1$ & $85.2 \pm 6.1$ & $96.3 \pm 10.5$ & $94.5 \pm 11.1$ & $82.7 \pm 8.7$ & $62.7 \pm 13.9$ \\
\hline
Fold-2 & $97.8 \pm 1.1$ & $88.9 \pm 4.9$ & $95.5 \pm 1.7$ & $86.0 \pm 5.1$ & $97.4 \pm 1.4$ & $95.5 \pm 4.1$ & $83.0 \pm 6.1$ & $61.5 \pm 13.9$ \\
\hline
Fold-3 & $97.8 \pm 0.9$ & $89.0 \pm 4.4$ & $94.3 \pm 6.3$ & $84.8 \pm 7.5$ & $97.5 \pm 2.3$ & $95.6 \pm 4.4$ & $81.5 \pm 9.4$ & $59.8 \pm 15.6$ \\
\hline
Fold-4 & $97.9 \pm 0.7$ & $89.2 \pm 3.9$ & $95.4 \pm 2.7$ & $86.2 \pm 5.4$ & $97.8 \pm 0.6$ & $96.2 \pm 2.7$ & $82.0 \pm 9.5$ & $59.3 \pm 16.7$ \\
\hline
Average & \cellcolor{lightred} $97.8 \pm 1.3$ & \cellcolor{lightred} $88.9 \pm 5.0$ & \cellcolor{lightgreen} $95.0 \pm 3.9$ & \cellcolor{lightgreen} $85.6 \pm 6.1$ & \cellcolor{lightblue} $97.2 \pm 5.5$ & \cellcolor{lightblue} $95.4 \pm 6.5$ & \cellcolor{lightyellow} $82.3 \pm 8.6$ & \cellcolor{lightyellow} $60.8 \pm 15.1$ \\
\hline
\end{tabular}
\end{table*}

\subsection{Testing protocols}

To optimize inference time as well as the required GPU memory, we implemented a minimal script for inference.
After loading an image, the same preprocessing strategy for the localization model as during training is performed before generating the model's prediction from which the \gls{roi} is computed.
Using the \gls{roi}, we also perform the same preprocessing strategy for the segmentation model as during training.
Lastly, we resample the prediction of our segmentation model such that it has the same size and spacing as the original image and copy all of the meta-data, i.e., origin, direction, orientation, etc., from it as well.

\section{Results}

We evaluate our segmentation model quantitatively as well as qualitatively in a 4-fold cross-validation.
Furthermore, we report the performance of our final submitted models on the hidden validation set, for which the organizers provided quantitative results as well as selected ground truth images after the challenge deadline.

\subsection{Quantitative Results of 4-fold Cross-Validation}

The evaluation of our 4-fold cross-validation confirms good performance of our segmentation model for multi-label segmentation, see Table~\ref{tab:cross-validation}.
The average \gls{dsc} of liver, kidney and spleen are $97.8\%$, $95.0\%$ and $97.2\%$, respectively, while the pancreas has the lowest \gls{dsc} of $82.3\%$.
In terms of \gls{nsd}, which is computed as the percentage of voxels for which the surface distance is below 1 mm, our method achieves a score of $88.9\%$, $85.6\%$ and $95.4\%$ for liver, kidney and spleen, respectively.
Again, the performance on the pancreas is the lowest, resulting in an \gls{nsd} of $60.8\%$.

\subsection{Quantitative Results on Validation Set}

Table~\ref{tab:quanti-validation} shows the performance of our submitted localization and segmentation models in terms of \gls{dsc} and \gls{nsd} on the hidden validation set.
Similar to the cross-validation results, the segmentations of liver, kidney and spleen were better by a large margin compared to the pancreas.
In terms of \gls{dsc}, liver, kidney and spleen achieved $95.43\%$, $89.74\%$ and $93.88\%$ on average, respectively, while the pancreas scored $75.15\%$.
The spleen performed best in terms of \gls{nsd} with $86.57\%$ followed by liver and kidney with $80.03\%$ and $77.76\%$ respectively.
Lastly, the pancreas achieved an average \gls{nsd} of $60.00\%$, meaning that $60.00\%$ of all voxels labelled as pancreas had a surface distance below 1 mm.

\begin{table}[!htbp]
\caption{Quantitative results on validation set.}
\label{tab:quanti-validation}
\centering
\begin{tabular}{ccc}
\hline
Organ    & DSC (\%)       & NSD (\%)        \\
\hline
\cellcolor{lightred} Liver    & \cellcolor{lightred} 95.43 $\pm$ 4.77 & \cellcolor{lightred} 80.03 $\pm$ 12.24 \\
\cellcolor{lightgreen} Kidney   & \cellcolor{lightgreen} 89.74 $\pm$ 13.71 & \cellcolor{lightgreen} 77.76 $\pm$ 15.89 \\
\cellcolor{lightblue} Spleen   & \cellcolor{lightblue} 93.88 $\pm$ 14.02 & \cellcolor{lightblue} 86.57 $\pm$ 17.55 \\
\cellcolor{lightyellow} Pancreas & \cellcolor{lightyellow} 75.15 $\pm$ 18.52 & \cellcolor{lightyellow} 60.00 $\pm$ 16.82 \\
\hline
\end{tabular}
\end{table}

\subsection{Qualitative Results}

\begin{figure}[htbp]
\centering
\includegraphics[width=0.47\textwidth]{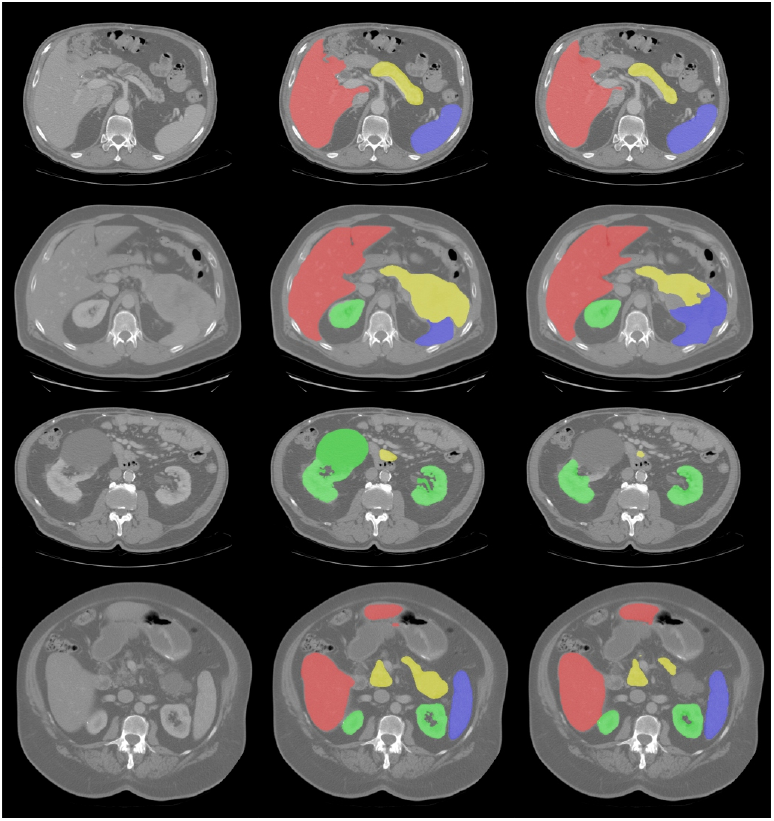}
\caption{Qualitative results of our 4-fold cross-validation showing the image (column 1), the ground truth (column 2) and the prediction of our model (column 3). A case of a successful prediction is given in row 1, while rows 2--4 represent difficult cases selected according to their low DSC.}
\label{fig:qualitative_result}
\end{figure}

\begin{figure}[htbp]
\centering
\includegraphics[width=0.47\textwidth]{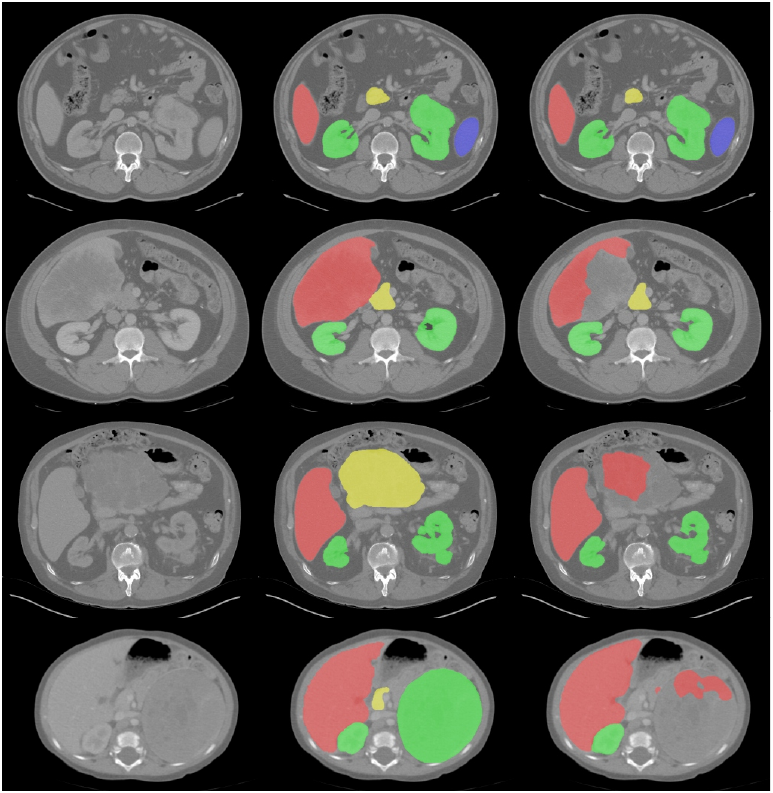}
\caption{Qualitative results of our submitted segmentation model on the validation set showing the image (column 1), the ground truth (column 2) and the model prediction (column 3). A case of a successful prediction is given in row 1, while rows 2--4 represent difficult cases selected by the challenge organizers.}
\label{fig:qualitative_results_validation_set}
\end{figure}

Selected qualitative results of our proposed model trained in the cross-validation are given in Fig.~\ref{fig:qualitative_result}, while Fig.~\ref{fig:qualitative_results_validation_set} shows results of our submitted model evaluated on the hidden validation set.
In both figures, the image, the ground truth label as well as the model's prediction are shown from left to right.
The segmented organs are the liver (red), the kidneys (green), the spleen (blue) and the pancreas (yellow).
The image given in the first row in Fig.~\ref{fig:qualitative_result} and~\ref{fig:qualitative_results_validation_set} represents a selected case, where the prediction of our model is very close to the ground truth for all labels.
In rows 2--4 in Fig.~\ref{fig:qualitative_result}, some of the most difficult cases from the cross-validation set are shown, for which we selected images with the lowest \gls{dsc}.
Similarly, rows 2--4 in Fig.~\ref{fig:qualitative_results_validation_set} show difficult cases of the validation set, which were selected by the challenge organizers.

\section{Discussion and Conclusion}

Observation of the quantitative results in Table~\ref{tab:cross-validation} shows that the segmentations of liver, kidney and spleen perform very well, each achieving a \gls{dsc} of at least $95\%$, while the segmentation of the pancreas only reached $82.3\%$.
The results of the \gls{nsd} show a similar trend, where again the segmentation of the pancreas performed worse by a large margin reaching only $60.8\%$, whereas the segmentations of the liver and kidney achieved more than $85\%$ on average.
The spleen showed the best results in \gls{nsd} with $95.4\%$, which means that less than $5\%$ of all voxels labelled as spleen had a surface distance of more than 1 mm.
The quantitative results of the submitted model on the validation set given in Table~\ref{tab:quanti-validation} follow the same trend, with liver, kidney and spleen performing much better in terms of \gls{dsc} and \gls{nsd} on average compared to the pancreas.
The segmentations of both liver and spleen achieved a \gls{dsc} of over $90\%$, closely followed by the kidney with $89.74\%$; however, the pancreas only managed to reach a \gls{dsc} of $75.15\%$.
In terms of \gls{nsd}, liver and spleen resulted in a score of over $80\%$ on the validation set, while the kidney reached $77.76\%$.
Again, the pancreas in comparison to the other labels underperformed by a large margin, achieving $60.00\%$ measured in \gls{nsd}.

When comparing the quantitative results of the cross-validation in Table~\ref{tab:cross-validation} to the results of the submitted model evaluated on the validation set in Table~\ref{tab:quanti-validation}, it can be observed that both \gls{dsc} and \gls{nsd} are lower on the validation set for all labels.
This gap in performance is caused by the data used to train and evaluate the respective model, as the cross-validation models were trained and evaluated only on subsets of the training set, while the submitted model was trained on the full training set and evaluated on the validation set.
As can be seen in Table~\ref{tab:dataset}, the training set only consists of images obtained from two centers, all acquired in the same phase.
In contrast, the validation set contains images from 9 different centers, 8 of which were not seen during training, as well as images acquired in various phases, many of which were also not seen during training.
Consequently, while the cross-validation can be used to evaluate how well the model performs on the task, the evaluation on the validation set shows how well the model generalizes to data from different centers with various phases.

When observing the qualitative results of difficult cases of the cross-validation set in Fig.~\ref{fig:qualitative_result}, rows 2--4, more closely, it becomes apparent that all three images are pathological and contain a rather large lesion that our model failed to label correctly.
More precisely, the image in row 2 contains a large lesion of the pancreas (yellow) of which some parts were labelled correctly by our model, while other parts of the lesion were either not labelled or even mislabelled as the spleen (blue).
In row 3, again, a very large lesion of the kidney (green) on the left can be observed, which is completely missing in the prediction of our model.
The image in the last row contains a lesion of the pancreas (yellow), which is also not labelled by our segmentation model.
Nevertheless, when looking at row 1, the labels of the healthy organs in rows 2--4 as well as the healthy parts of the organs containing these huge lesions, it can be observed that the predictions are very close to the ground truth labels.
Similarly, the qualitative results of difficult cases of the validation set in Fig.~\ref{fig:qualitative_results_validation_set} given in rows 2--4 contain very large lesions that were not labelled correctly by our model.
In row 2, the liver (red) contains a large lesion spanning almost the entire liver in this slice.
The spleen (blue) in row 3 as well as the right kidney (green) in row 4 of the presented slices also contain lesions that lead to a huge enlargement of the respective organ.
All three of these huge lesions could not be labelled correctly by our segmentation model and in some cases even led to our model mislabelling these regions.
Nevertheless, healthy and less severely pathological organs were segmented very well, including small- and medium-sized lesions as can be observed from the right kidney (green) in row 3, bottom, and again the right kidney in row 1, top.

A closer look at the dataset reveals that many of the organs per image are either healthy or only contain smaller lesions, which are easier to learn and thus more likely to be segmented correctly.
In contrast, large lesions of individual organs are only present in a small number of images, which can lead to poor segmentation predictions for unseen images if this type of pathology was never or not sufficiently encountered during training.
This indicates that additional data containing very large lesions would be beneficial to solve the problems observed in these regions.
Other potential solutions, which are not based on acquiring additional data, are to either use an imbalanced training scheme, where images with large lesions are selected more frequently during training, or a more sophisticated data augmentation strategy, where large lesions are injected artificially into the image using, e.g., a local inflation operation.

\section*{Acknowledgment}
The authors of this paper declare that the segmentation method they implemented for participation in the FLARE challenge has not used any pre-trained models or additional datasets other than those provided by the organizers.
This work was supported by the Austrian Research Promotion Agency (FFG): 871262. 

{\small
\bibliographystyle{IEEEtran} 
\bibliography{main}
}

\end{document}